\documentclass[sigconf]{acmart}

\AtBeginDocument{%
  \providecommand\BibTeX{{%
    \normalfont B\kern-0.5em{\scshape i\kern-0.25em b}\kern-0.8em\TeX}}}

\copyrightyear{2023} 
\acmYear{2023} 
\setcopyright{acmlicensed}\acmConference[SIGIR '23]{Proceedings of the 46th International ACM SIGIR Conference on Research and Development in Information Retrieval}{July 23--27, 2023}{Taipei, Taiwan}
\acmBooktitle{Proceedings of the 46th International ACM SIGIR Conference on Research and Development in Information Retrieval (SIGIR '23), July 23--27, 2023, Taipei, Taiwan}
\acmPrice{15.00}
\acmDOI{10.1145/3539618.3591641}
\acmISBN{978-1-4503-9408-6/23/07}

\usepackage{latexsym}

\usepackage{microtype}

\usepackage{booktabs}
\usepackage{amsmath}
\usepackage{algorithm}
\usepackage{algpseudocode}
\usepackage{multirow}
\usepackage{mathtools}
\usepackage{subfigure}
\usepackage{graphicx}
\usepackage{balance}
\usepackage{color}
\usepackage{enumitem}
\usepackage{xspace}

\renewcommand{\vec}[1]{\ensuremath{\mathbf{#1}}}
\newcommand{\stitle}[1]{\vspace{1mm} \noindent {\bf #1}}
\newcommand{\eg}{{\it e.g.}}

\newcommand{\ie}{{\it i.e.}}

\newcommand{\bL}{\ensuremath{\mathcal{L}}}
\newcommand{\bD}{\ensuremath{\mathcal{D}}}

\newcommand{\bS}{\ensuremath{\mathcal{S}}}
\newcommand{\bQ}{\ensuremath{\mathcal{Q}}}

\newcommand{\bG}{\ensuremath{\mathcal{G}}}
\newcommand{\bN}{\ensuremath{\mathcal{N}}}
\newcommand{\bE}{\ensuremath{\mathcal{E}}}

\newcommand{\model}{G2P2}

\begin{document}

\title{Augmenting Low-Resource Text Classification with Graph-Grounded Pre-training and Prompting}

\author{Zhihao Wen}
\affiliation{%
  \institution{Singapore Management University}
   \country{Singapore}
   }
\email{zhwen.2019@smu.edu.sg}

\author{Yuan Fang}
\affiliation{%
  \institution{Singapore Management University}
  \country{Singapore}
  }
\email{yfang@smu.edu.sg}

\renewcommand{\shortauthors}{Zhihao Wen and Yuan Fang}

\begin{abstract}
Text classification is a fundamental problem in information retrieval with many real-world applications, such as predicting the topics of online articles and the categories of e-commerce product descriptions.
However, low-resource text classification, with few or no labeled samples, poses a serious concern for supervised learning.
Meanwhile, many text data are inherently grounded on a network structure, such as a hyperlink/citation network for online articles, and a user-item purchase network for e-commerce products.
These graph structures capture rich semantic relationships, which can potentially augment low-resource text classification. 
In this paper, we propose a novel model called Graph-Grounded Pre-training and Prompting (\model) to address low-resource text classification in a two-pronged approach. During pre-training, we propose three graph interaction-based contrastive strategies to jointly pre-train a graph-text model; during downstream classification, we explore 
prompting
for the jointly pre-trained model to achieve low-resource classification. Extensive experiments on four real-world datasets demonstrate the strength of \model\ in zero- and few-shot low-resource text classification tasks. 

\end{abstract}

\ccsdesc[500]{Information systems~Content analysis and feature selection}
\ccsdesc[500]{Information systems~Clustering and classification}

\keywords{Text classification, graph neural networks, low-resource learning, pre-training, prompt-tuning}

\maketitle

\section{Introduction}
Text classification is a fundamental research problem with many important applications in information retrieval. For example, predicting the topics of online articles can help readers easily search and navigate within the website or portal \cite{mccallum2000automating}, and classifying the category of e-commerce item descriptions enables businesses to structure their inventory efficiently and improve users' search experience \cite{xu2019open}.
Advances in supervised deep learning in the last decade have achieved remarkable success for text classification, especially when there are large-scale and high-quality labeled data. However, data labeling is often costly and time-consuming, making low-resource classification, in which no or few labeled samples are available, an appealing alternative. 


To address low-resource text classification, one approach is to utilize pre-trained language models (PLMs) \cite{kenton2019bert, radford2018improving}, many of which are based on the transformer architecture \cite{vaswani2017attention} due to its powerful ability of encoding texts. 
A PLM can be adapted to different tasks by \emph{fine-tuning} the model parameters to task-specific objectives. While the ``pre-train, fine-tune'' paradigm requires fewer labeled data than traditional supervised learning, it suffers from two drawbacks. First, state-of-the-art PLMs typically have a huge model size, \eg,  GPT-3 has 175 billion parameters \cite{brown2020language}, making fine-tuning prohibitively expensive \cite{li2021prefix}. Second, fine-tuning still needs a reasonable amount of labeled data due to the gap between pre-training and fine-tuning objectives, and thus struggles with low-resource scenarios including zero- and few-shot classification.  
%
To overcome the problem of pre-training and fine-tuning, \emph{prompting} \cite{brown2020language} has been proposed. It uses a natural language instruction or ``prompt'' to give a hint of the downstream task, whilst freezing the parameters of a large PLM. In other words, no fine-tuning or additional training is required at all for a new task. 
However, discrete natural language prompts can be difficult to design and may result in suboptimal performance compared to fine-tuning \cite{lester2021power}. More recently,
\emph{prompt tuning} \cite{liu2021gpt, lester2021power} formulates a continuous prompt as a learnable embedding, which is optimized during task adaptation without updating the PLM. 

Meanwhile, text data are frequently grounded on network structures, such as hyperlink or citation networks for online articles, and user-item interaction graphs for e-commerce. These graph structures expose valuable relationships between articles or items, which can be used to augment low-resource text classification. While existing PLMs and prompting do not exploit these relationships, graph neural networks (GNNs) \cite{wu2020comprehensive} are designed to learn from graph structures based on a message-passing architecture.
However, traditional end-to-end training of GNNs heavily relies on abundant task-specific labels, which motivates self-supervised GNNs \cite{wu2021self} that employ well-designed pretext tasks on a label-free graph \cite{velickovic2019deep,hu2019strategies,hu2020gpt}.
Unfortunately, the treatment of text features in GNNs remains rudimentary. Typically, a simple bag-of-words representation \cite{yang2016revisiting} or aggregation of shallow word embedding vectors \cite{mikolov2013efficient} is fed into GNNs as the initial node features, which are further propagated along graph structures. 
Hence, the modeling of texts in GNNs is coarse-grained, unable to fully capture the subtle semantic differences and similarities within texts. 


\stitle{Challenges and present work.}
To overcome the limitations of existing text- and graph-based solutions, we must address two open questions as follows.


Firstly, \emph{how do we capture fine-grained textual semantics, while leveraging graph structure information jointly?}
A na\"ive approach is to use a language model to generate features from raw texts as input, and then train a GNN. However, in this way, the texts and graph are only loosely coupled, lacking an explicit pairing to complement each other.  
In this paper, we propose graph-grounded contrastive pre-training, to maximize the alignment between text and graph representations based on three types of graph interaction, namely, text-node, text-summary, and node-summary interactions.

Secondly, \emph{how do we augment low-resource text classification given a jointly pre-trained graph-text model?} 
We propose a novel approach of ``prompting'' a jointly pre-trained graph-text model instead of fine-tuning it. This allows us to leverage the most relevant structural and semantic information from the pre-trained model, making the process friendlier to low-resource scenarios. More specifically, we use handcrafted discrete prompts for zero-shot classification, and continuous prompts for few-shot settings based on automatic prompt-tuning. Due to the significantly fewer parameters involved, prompt-tuning is more label- and computation-efficient than fine-tuning the pre-trained model. Furthermore, we propose a context-based initialization for prompt-tuning that considers graph structures between texts to provide a more informative starting point.

%

\stitle{Contributions.}
To summarize, we make the following contributions in this work. 
    (1) This is the first attempt to pre-train text and graph encoders jointly for low-resource text classification.
    (2) We propose a novel model called Graph-Grounded Pre-training and Prompting (\model), with three graph interaction-based constrastive strategies in pre-training, and a prompting approach for the jointly pre-trained graph-text model in downstream tasks.
    (3) We conduct extensive experiments on four real-world datasets to demonstrate the strength of \model\ in zero- and few-shot text classification.

\begin{figure*}[t]
  \centering
  \includegraphics[width=0.99\linewidth]{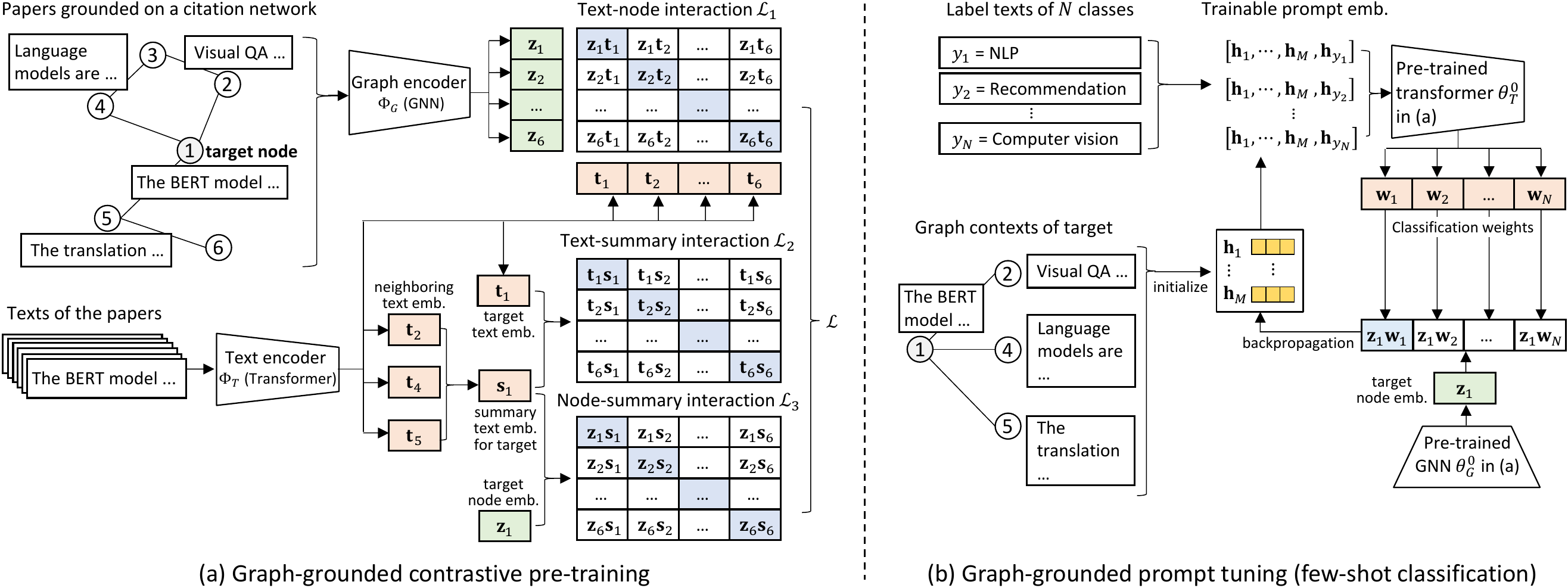}
  \caption{Overall framework of \model. (a) During pre-training, it jointly trains a text and a graph encoder through three contrastive strategies. (b) During testing, it performs prompt-assisted zero- or few-shot classification (the figure only shows prompt-tuning for few-shot classification, while zero-shot inference adopts a simplified scheme).}
  \label{fig:framework}
\end{figure*}

\section{Related Work}

\stitle{Graph neural networks.}
Inspired by the success of convolutional networks in computer vision, GNNs have emerged to handle non-Euclidean relational data \cite{wu2020comprehensive}, ranging from early semi-supervised models such as GCN \cite{KipfW17}, GAT \cite{velivckovic2018graph} and GIN \cite{XuHLJ19}, to the more recent self-supervised pre-training paradigm \cite{velickovic2019deep, hu2019strategies, hu2020gpt, lu2021learning}.
Besides their widespread success in graph tasks, they have also been leveraged to improve text-based tasks through knowledge graphs \cite{cao2021dekr} and heterogeneous graphs \cite{linmei2019heterogeneous}, or multi-modal learning \cite{liu2021mm}. 
However, these approaches either employ coarse-grained text treatment, or have decoupled graph and text encoders without fully exploiting the intrinsic relationship between them. Although a more recent approach called GLEM \cite{zhao2023learning} integrates both the text and graph structure information by fusing language models and GNNs, it is not designed for low-resource learning.

\stitle{Language pre-training and prompting.}
Pre-trained language models \cite{han2021pre} have become the most popular backbone in natural langauge processing (NLP). While earlier PLMs such as GPT \cite{radford2018improving}, BERT \cite{kenton2019bert}, XLNet \cite{yang2019xlnet} and RoBERTa \cite{liu2019roberta} still have affordable model size, recent introductions such as T5 \cite{raffel2020exploring} and GPT-3 \cite{brown2020language} produce massive models with billions of parameters.
To avoid the high cost of fine-tuning on these large models, prompting \cite{liu2021pre} starts to receive more attention in the community.
A prompt is a special template to pad the task input, with the goal of extracting useful knowledge from PLMs to flexibly adapt to downstream tasks.
Fueled by the success of GPT-3, numerous prompting methods including discrete natural language prompt \cite{shin2020autoprompt, schick2021exploiting, gao2021making} and continuous prompt \cite{li2021prefix, lester2021power, liu2021gpt, qin2021learning,  zhong2021factual} have emerged. The strength of prompting has been validated in a wide range of NLP applications, including text classification \cite{hu2022knowledgeable, min2022noisy,sun2021nsp, zhang2021aspect, han2021ptr}, machine translation \cite{tan2022msp} and relation extraction \cite{chen2022knowprompt, sainz2021label}. More recently, prompting has also been applied to GNNs for node classification \cite{sun2022gppt}.

\stitle{Zero- or few-shot paradigms.}
Broadly speaking, our setting is also related to other learning paradigms. For example, in semi-supervised learning \cite{MiyatoDG17, xie2020unsupervised, chen2020mixtext}, each class may only have a few examples, but all classes must be seen in training and they cannot handle any novel class during testing.
Meta-learning \cite{finn2017model, yu2018diverse, han2018fewrel, bansal2020self, bao2020few, zhou2019meta, wang2020graph, wen2021meta, wen2023generalizing} is another popular paradigm that supports few-shot learning. However, large-scale labeled data are still required in a so-called ``meta-training'' phase, to support the few-shot learning of novel classes during ``meta-testing''.
In contrast, we only need label-free data for pre-training, without requiring any meta-training phase that would consume large-scale labeled data. 
Separately, there also exists joint consideration of image and text data using a contrastive pre-training strategy for zero- or few-shot classification
 \cite{radford2021learning}. In our work, graph data are significantly different from images, which provide various types of interaction between texts. On graphs, zero-shot node classification has also been done \cite{wang2021zero}. It relies heavily on the availability of Wikipedia pages or other side information to generate class prototype embeddings. However, it is very labor intensive to find and curate the right side information, especially when there are a large number of classes and/or novel classes emerge frequently.


\section{Proposed Approach} 
In this section, we introduce our approach \model\ for low-resource text classification. We start with some preliminaries and an overview, and then present the details of the proposed approach.

\subsection{Preliminaries} 


\stitle{Graph-grounded text corpus.} Consider a set of documents $\bD$, which is grounded on a graph $\bG=(\bD, \bE, \vec{X})$ such that each document $d_i\in\bD$ is a node $v_i$ in the graph. The documents are linked via edges in $\bE$, which are formed based on the application (\eg, if each document represents an article, the edges could be citations between articles). Each node $v_i$ is also associated with a feature vector $\vec{x}_i$, given by the input feature matrix $\vec{X}$. Finally, each document/node\footnote{We will use ``node'' and ``document'' interchangeably given their one-one correspondence in our context.} has a class label (\eg, the topic of the article). 


\stitle{Low-resource classification.}
A low-resource task consists of a support set $\bS$ and a query set $\bQ$. The support set $\bS$ contains $N$ classes, and each class has $K$ labeled examples where $K$ is a small number (\eg, 1 or 5), known as $N$-way $K$-shot classification. The query set $\bQ$ contains one or more unlabeled instances belonging to the $N$ classes in the support set. Our goal is to classify the instances in the query set based on the labeled examples in the support set.
Unlike episodic few-shot meta-learning \cite{finn2017model} which has both training tasks and testing tasks, we only have testing tasks; in the training stage, we perform self-supervised pre-training on label-free data only. 
As a special case, tasks with $K=0$ are known as zero-shot classification, which means that there is no labeled example at all and we can only rely on class metadata (\eg, class label text).



\subsection{Overview of \model}

As shown in Fig.~\ref{fig:framework}, our model consists of two stages: (a) graph-grounded constrastive pre-training, and (b) graph-grounded prompt-tuning for low-resource classification. 
 
During pre-training,  we learn a dual-modal embedding space by jointly training a text encoder and graph encoder in a self-supervised fashion, since a document also exists as a node on the graph. More specifically, we use a transformer-based text encoder and a GNN-based graph encoder. The transformer takes the text on each node (\ie, document) as the input, and outputs a text embedding vector $\vec{t}_{i}$ for node $v_i$. On the other hand, the GNN takes the graph and node features as input, and generates a node embedding vector $\vec{z}_i$ for node $v_i$. Subsequently, in the dual-modal embedding space, we align the text and graph representations on the same or related nodes through three contrastive strategies based on different types of interaction on the graph. 

In downstream testing, we employ prompting on our jointly pre-trained graph-text model for zero- or few-shot classification. For zero-shot classification, we use handcrafted discrete prompts together with the label text. For few-shot classification, we use continuous prompts to pad the label text. In particular, for prompt-tuning, we initialize the continuous prompt embeddings based on graph contexts. 

\subsection{Graph-grounded contrastive pre-training}

The graph-grounded pre-training learns a dual-modal embedding space by jointly training a text encoder and a graph encoder, based on three types of interaction on the underlying graph.

\stitle{Dual encoders.} 
The text encoder is a transformer \cite{vaswani2017attention}, which we denote $\Phi_T$. Given a document $d_i$, the text encoder\footnote{Technically, the input to the text encoder is a sequence of continuous embeddings; the tokens in a document are first converted to word embeddings.} outputs the $d$-dimensional embedding vector of $d_i$, denoted $\vec{t}_i\in \mathbb{R}^d$:
\begin{align}
\label{eq:text emb}
    \vec{t}_i = \Phi_T(d_i;\theta_T),
\end{align}
where $\theta_T$ represents the parameter set of the transformer.
Correspondingly, let $\vec{T} \in \mathbb{R}^{|\bD|\times d}$ represent the text embedding matrix for all documents.

At the same time, a document $d_i$ is also a node $v_i$ in the graph. We choose a classic GNN called graph convolutional network (GCN) \cite{KipfW17} as the graph encoder, denoted $\Phi_Z$. It similarly outputs an embedding vector $\vec{z}_i\in \mathbb{R}^d$ for a given node $v_i$:
\begin{align}
\label{eq:node emb}
    \vec{z}_i= \Phi_Z(v_i;\theta_G), 
\end{align}
where $\theta_G$ represents the parameter set of the GCN.
Likewise, let $\vec{Z} \in \mathbb{R}^{|\bD|\times d}$ represent the graph embedding matrix for all nodes.

\stitle{Text-node interaction.}
\label{sec:model:node level}
Our graph-grounded texts naturally implies a bijection between nodes and texts, where each document $d_i$ corresponds to the node $v_i$ in the graph.
Inspired by the pairing of image and its caption text \cite{radford2021learning} and the mapping of content and node sequences \cite{liu2018content}, we design a pre-training strategy to predict which text document matches which node in the graph.


Specifically, given $n$ documents and the corresponding $n$ nodes, there are $n^2$ possible document-node pairs $\{(d_i,v_j)\mid i,j =1,\ldots,n \}$.
Among them, only $n$ pairs with $i=j$ are true matching, whereas the remaining $n^2-n$ pairs are false matching. 
%
%
As our first contrastive strategy, we exploit the bijective interaction between texts and nodes on the graph, to maximize the cosine similarity of the $n$ matching pairs, while minimizing the cosine similarity of the $n^{2}-n$ unmatching pairs. To compute the cosine similarity for the $n^2$ pairs, we first perform a row-wise L2 normalization on embedding matrices $\vec{T}$ and $\vec{Z}$ to obtain $\tilde{\vec{T}}$ and $\tilde{\vec{Z}}$, respectively. We then compute a node-text similarity matrix $\vec{\Lambda}_1 \in \mathbb{R}^{n\times n}$ to capture pairwise cosine similarity, as follows.
%
\begin{align}
\label{eq:sm1}
    \vec{\Lambda}_{1}= \left( \tilde{\vec{Z}} \tilde{\vec{T}}^{\top} \right) \cdot \exp(\tau), 
\end{align}
where $\tau\in \mathbb{R}$ is a trainable temperature parameter to scale the similarity values \cite{radford2021learning}. 

\textsc{\textbf{Remark.}} Although $\vec{\Lambda}_{1}\in\mathbb{R}^{n\times n}$ is a dense matrix, it is constructed batchwise for practical implementation. That is, $n$ is not the total number of documents, but the relatively small batch size, and thus the overhead is negligible. $\vec{\Lambda}_{2}$ and $\vec{\Lambda}_{3}$ will be introduced later following the same treatment. $\hspace*{\fill}\qed$

To formulate the contrastive loss based on the text-node bijective interaction, we adapt the \emph{multi-class N-pair loss} \cite{sohn2016improved,zhang2020contrastive}, by considering both the row-wise and column-wise cross entropy loss w.r.t.~the row or column index. For example, the $i$-th row of $\vec{\Lambda}_1$ represents the similarity scores between node $v_i$ and every document, in which the row index $i$ indicates the ground truth document $d_i$ that matches $v_i$.
\begin{align}
\label{eq:node level}
    \bL_1 = \textstyle\frac{1}{2}\left(\textsc{CE}(\vec{\Lambda}_1, \vec{y}) + \textsc{CE}(\vec{\Lambda}_1^{\top}, \vec{y}) \right),
\end{align}
where $\vec{y}=(1,2,\ldots,n)^\top$ is the label vector for contrastive training, and \textsc{CE} denotes the cross entropy loss applied to the input matrix $\vec{\Lambda}_1$ or $\vec{\Lambda}_1^\top$ in a row-wise manner.

\stitle{Text-summary interaction.}
\label{sec:model:TT}
Apart from the bijective text-node interaction, we further exploit  higher-order interactions on the graph. 
In particular, each document has a set of neighboring documents defined by graph topology. The neighboring documents can be understood as a summary of the target document given the semantic relatedness between them. For example, on an e-commerce network, the products purchased by a user naturally portray a summary of the user and vice versa. 
%
Without loss of generality, we employ a simple mean pooling to generate the summary embedding $\vec{s}_i\in \mathbb{R}^d$ as follows.
\begin{align}
\label{eq:context emb}
    \vec{s}_{i}=\textstyle\frac{1}{|\bN_i|} \sum_{j \in \bN_{i}} \vec{t}_{j}.
\end{align}
For efficiency, we only sample a fixed number of neighboring documents to generate the summary. 
Then, let $\vec{S} \in \mathbb{R}^{n\times d}$ denote the summary text embedding matrix for all documents.

Hence, as our second contrastive strategy, we seek to align the text embedding of each document and its corresponding summary text embedding, based on the text-summary interaction derived from graph neighborhood. In other words, we maximize the cosine similarity of the $n$ matching pairs of document and its neighborhood-based summary, while minimizing the cosine similarity of the $n^{2}-n$ unmatching pairs. 
Specifically, we first follow Eq.~\eqref{eq:sm1} to construct a text-summary similarity matrix  $\vec{\Lambda}_2 \in \mathbb{R}^{n\times n}$:
\begin{align}
\label{eq:sm2}
    \vec{\Lambda}_{2}= \left( \tilde{\vec{T}} \tilde{\vec{S}}^{\top} \right) \cdot \exp(\tau).
\end{align}
Subsequently, we apply the same contrastive loss following Eq.~\eqref{eq:node level}, as follows.
%
\begin{align}
\label{eq:TT}
    \bL_2 = \textstyle\frac{1}{2}\left(\textsc{CE}(\vec{\Lambda}_2, \vec{y}) + \textsc{CE}(\vec{\Lambda}_2^{\top}, \vec{y}) \right),
\end{align}

\stitle{Node-summary interaction.}
The neighborhood-based summary for document $d_i$ also serves as a semantic description of node $v_i$.
Mirroring the text-summary interaction, as our third contrastive strategy, we seek to align the node embedding and its neighborhood-based summary text embedding. In the following, we similarly compute a node-summary similarity matrix  $\vec{\Lambda}_3 \in \mathbb{R}^{n\times n}$, and formulate the corresponding contrastive loss $\bL_3$.
\begin{align}
    \vec{\Lambda}_{3}&= \left( \tilde{\vec{Z}} \tilde{\vec{S}}^{\top} \right) \cdot \exp(\tau),\label{eq:sm3}\\
    \bL_3 &= \textstyle\frac{1}{2}\left(\textsc{CE}(\vec{\Lambda}_3, \vec{y}) + \textsc{CE}(\vec{\Lambda}_3^{\top}, \vec{y}) \right).\label{eq:ZT}
\end{align}


\stitle{Overall pre-training objective.}
Finally, we integrate the three contrastive losses based on the text-node, text-summary and node-summary interactions. 
We obtain a pre-trained model $\theta^0=(\theta_{T}^0, \theta_{G}^0)$ consisting of the parameters of the dual encoders, given by 
\begin{align}
\label{eq:final loss}
    \theta^0 = \arg\min_{\theta_{T}, \theta_{G}} \bL_{1} + \lambda (\bL_{2} + \bL_{3}),
\end{align}
where $\lambda \in \mathbb{R}^+$ is a hyperparameter to balance the contribution from summary-based interactions. 

The pre-training procedure is outlined in Algorithm~\ref{alg:pre-train}, which has the following complexity per epoch. Let $|\bD|$ be the number of documents, $\eta$ be the number of neighbors sampled to generate the summary embedding in Eq.~\eqref{eq:context emb}, and $\beta$ be the batch size. First, the cost of generating the three types of embeddings (lines 5--8) per epoch is  $O(|\bD|\eta)$, given that calculating the summary embedding needs go through $\eta$ neighbors.
Second, the cost of calculating the three similarity matrices in each batch is $O(\beta^2)$, and the total cost per epoch is $O\left(\frac{|\bD|}{\beta}\beta^2\right)=O(|\bD|{\beta})$ given $\frac{|\bD|}{\beta}$ batches in an epoch.
Thus, the overall complexity is $O(|\bD|(\eta+\beta))$, which is linear in the number of documents, since $\eta$ and $\beta$ are small constants. In our implementation, we set $\eta=3$ and $\beta=64$.
\begin{algorithm}[t]
\small
\caption{\textsc{Pre-training Procedure of \model}}
\label{alg:pre-train}
\begin{algorithmic}[1]
    \Require A graph-grounded text corpus $\bG=(\bD, \bE, \vec{X})$.
    \Ensure Pre-trained weights of text encoder $\theta^0_{T}$, graph encoder $\theta^0_{G}$.
    \State $\theta^0_{T},\theta^0_{G} \gets$ parameters initialization;
    \While{not converged}
        \State sample batches of documents from $\bD$;
        \For{each batch}
            \For{each node $v_{i}$/document $d_{i}$ in the batch} 
                \State calculate $d_{i}$'s text embedding $\vec{t}_{i}$; \Comment{Eq.~\eqref{eq:text emb}}
                \State calculate $v_{i}$'s node embedding $\vec{z}_{i}$;
                \Comment{Eq.~\eqref{eq:node emb}}
                \State calculate $v_{i}$'s summary embedding $\vec{s}_{i}$;
                \Comment{Eq.~\eqref{eq:context emb}}
            \EndFor 
            \State calculate the similatity matrices $\vec{\Lambda}_{1},\vec{\Lambda}_{2},\vec{\Lambda}_{3}$;
            \Comment{Eqs.~\eqref{eq:sm1}, \eqref{eq:sm2}, \eqref{eq:sm3}}
            \State calculate the contrastive losses $\bL_{1}$, $\bL_{2}$, $\bL_{3}$;
            \Comment{Eqs.~\eqref{eq:node level},~\eqref{eq:TT},~\eqref{eq:ZT}}
            \State update the overall loss $\bL$;           
            \Comment{Eq.~\eqref{eq:final loss}}
            \State $\theta^0_{T},\theta^0_{G}\gets$ update via backpropagation  
        \EndFor 
    \EndWhile 
    \State \Return $\theta^0_{T},\theta^0_{G}$.
\end{algorithmic}
\end{algorithm}

\subsection{Prompting joint graph-text model}\label{sec:model_prompt}
After pre-training our graph-text model, it is non-trivial to apply it to low-resource classification. To narrow the gap between pre-training and downstream tasks, the traditional ``pre-train, fine-tune'' paradigm typically introduces a new projection head for the downstream task, which will be fine-tuned together with the whole pre-trained model. However, in a low-resource setting, it is neither effective nor efficient to update the entire model with a huge number of parameters.
Without updating massive PLMs, prompting has recently emerged as a powerful alternative to fine-tuning in NLP  \cite{liu2021pre}. However, prompting has not been explored for graph-text models, where structural and textual information have been jointly pre-trained. In the following, we elaborate on our prompting strategies for zero- and few-shot classification.

\stitle{Zero-shot classification.}
In the zero-shot setting, we can only use handcrafted discrete prompts, as the absence of labeled data in zero-shot tasks cannot support learnable prompts.

\begin{figure}[t]
   \includegraphics[width=0.9\linewidth]{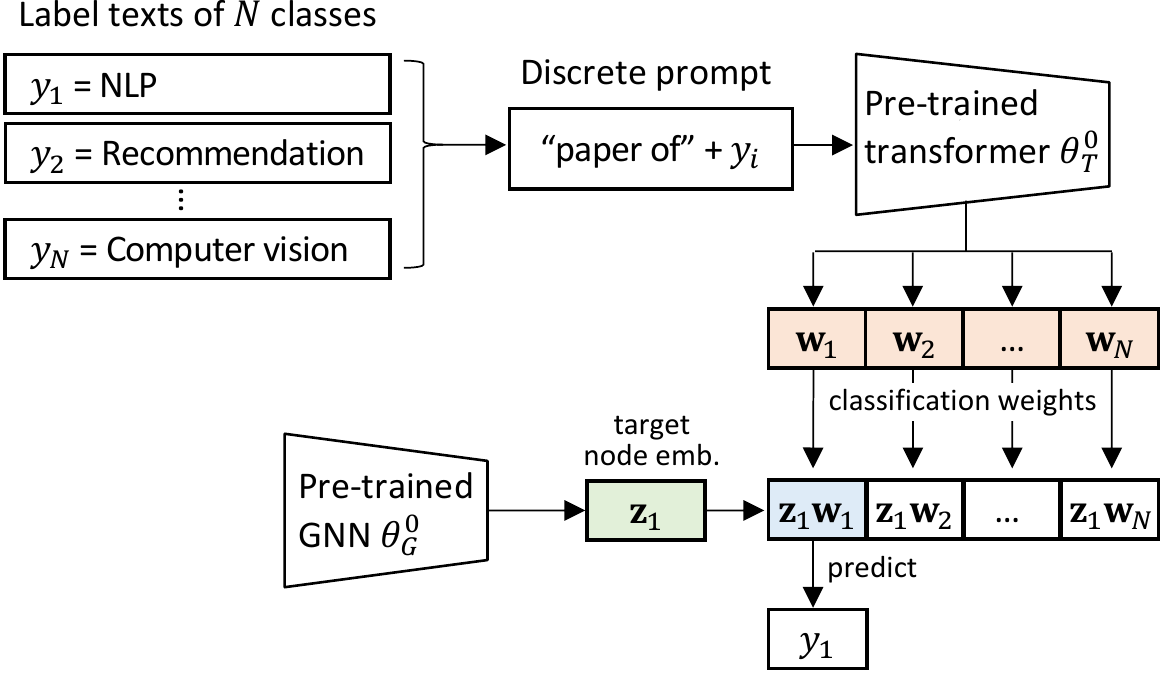}
	\caption{Schematic diagram for zero-shot classification. The pre-trained models $\theta_G^0$ and $\theta_T^0$ are obtained from Fig.~\ref{fig:framework}(a).}
	\label{fig:zero-shot}
\end{figure}

In $N$-way zero-shot classification, out of $N$ classes, we predict the class which has the highest similarity to the given node. As illustrated by the diagram in Fig.~\ref{fig:zero-shot}, the classification weights can be generated by the text encoder based on the class label texts \cite{wang2018joint}, without requiring any labeled sample for the classification task. Specifically, the weight vector $\vec{w}_y$ for class $y\in\{1,2,\ldots,N\}$ is the output of the pre-trained text encoder, \ie,
\begin{align}
    \vec{w}_y=\phi_T(\text{``\texttt{prompt [CLASS]}''}; \theta_T^0).
\end{align}
Here ``\texttt{prompt [CLASS]}'' is a prompt template, where \texttt{[CLASS]} refers to the label text of the target class $y$ (\eg, ``\texttt{NLP}'' for paper area classification), and \texttt{prompt} is a manually engineered sequence of natural language tokens to signal the relevance of the label text (\eg, ``\texttt{paper of NLP}'' helps focus on the topic of the paper). In the simplest case, ``\texttt{prompt}'' can be an empty string so that we only rely on the label text.
Then, the class distribution given node representation $\vec{z}_i$ is predicted as 
\begin{align}
    p(y \mid \vec{z}_i)=\frac{\exp \left(\langle\vec{z}_i,\vec{w}_{y}\rangle\right)}{\sum_{y=1}^{N} \exp \left(\langle\vec{z}_i, \vec{w}_{y}\rangle\right)},\label{eq:softmax_clf}
\end{align}
where $\langle\cdot, \cdot\rangle$ is the cosine similarity.

\stitle{Few-shot classification.}
The problem with discrete prompts is that they are difficult to optimize, 
given that  PLMs are intrinsically continuous.
Substituting discrete natural language prompts with learnable continuous prompts, prompt tuning \cite{lester2021power,liu2021gpt,liu-etal-2022-p} can automate the optimization of prompts when some labeled data are available. 
Hence, in the few-shot setting, we explore prompt tuning to cue in the relevant structural and semantic information from our jointly pre-trained graph-text model.

Specicifally, instead of a sequence of discrete tokens, we take a sequence of 
continuous embeddings   $[\vec{h}_{1}, \cdots, \vec{h}_{M}, \vec{h}_\texttt{CLASS}]$ as the prompt, 
where $M$ is a hyperparameter indicating the number of context tokens, each $\vec{h}_{m}$ ($m\le M$) is a trainable vector, 
and $\vec{h}_\texttt{CLASS}$ is the word embedding sequence of the target class label. 
The continuous prompt is fed as input to the text encoder to generate the classification weights for each class $y$:
\begin{align}
\label{eq:prompt embeddings}
    \vec{w}_y=\phi_T([\vec{h}_{1}, \cdots, \vec{h}_{M}, \vec{h}_\texttt{CLASS}]; \theta_T^0),
\end{align}
where each $\vec{h}_{m}$ ($m \le M$) has the same dimension as the input word embeddings to the text encoder. 
%

Using the same softmax layer in Eq.~\eqref{eq:softmax_clf}, 
we further update the continuous prompt embeddings using the labeled support set of the few-shot task by minimizing a cross entropy loss, whilst freezing the parameters of the dual encoders. This prompt tuning process is both data- and computation-efficient, given the small number of learnable parameters in the prompt. 
%



Furthermore, existing prompt tuning methods either initialize the prompt embeddings randomly \cite{lester2021power,liu-etal-2022-p} or using the word embeddings of handcrafted discrete prompts \cite{zhou2022learning}. While random initialization is non-informative and more prone to local optimum, it is still difficult to pick the right discrete prompts for initialization. 
 Therefore, we take the advantage of graph structures to initialize the prompt embeddings. 
 
 Specifically, given a node $v_i$, we define its \emph{graph contexts} as its neighbor set $\{v_j\mid j\in \bN_i\}$. Due to the underlying semantic relatedness, the graph contexts of the few-shot examples carry strong signals about the task, which can be exploited to improve the initialization.
For each document/node $v_i$ in the task support set, we sample $\eta$ nodes from its graph contexts. For $v_i$ itself and each context node sampled, we truncate its corresponding document to $M$ words, and convert it to a sequence of $M$ word embedding vectors, each having the same dimension as the vector $\vec{h}_m$ ($m\le M$) in our continuous prompt. Hence, for each support node, we would obtain $\eta+1$ such sequences; in an $N$-way $K$-shot task, there is 
a total of $NK(\eta+1)$ sequences.
We take the average of these embedding sequences to initialize the learnable prompt vectors $\vec{h}_1,\ldots,\vec{h}_M$, which is derived from graph contexts and thus could provide a more informative starting point than random initialization. 

\section{Experiments}
We conduct extensive experiments to evaluate our proposed approach \model
, with comparison to state-of-the-art baselines and model analyses.

\subsection{Experimental setup}
\stitle{Datasets.}
Four public graph-grounded text corpora are used, as summarized in Tab.~\ref{table.datasets}.
\begin{itemize}[leftmargin=*]
    \item \textbf{Cora}: 
    Known as the ``Cora Research Paper Classification'' dataset \cite{McCallumIRJ}, it is a collection of research papers that are linked to each other through citations. 
    The abstract of a paper is deemed a text document. The papers are classified into a topic hierarchy with 73 leaves. After removing papers with no content or label, the resulting hierarchy has 70 leaf topics. Note that we are using a more comprehensive version of the Cora dataset, which is larger and has more classes than the version used elsewhere \cite{KipfW17}.
    \item \textbf{Art}, \textbf{Industrial} and \textbf{Music Instruments (M.I.)} 
    are three Amazon review datasets \cite{ni2019justifying}, respectively from three broad areas, namely, arts, crafts and sewing (Art), industrial and scientific (Industrial), and musical instruments (M.I.). 
     The description of each product is deemed a text document, whereas the reviews of a user are combined into one document to reflect the user's preferences. If a user has reviewed a product, a link is constructed between them. The product subcategories within a broad area represent the classes, which are fine-grained and may involve thousands of classes with subtle differences. The classification is only performed on product descriptions, whereas the user reviews only serve to enrich the text semantics. 
\end{itemize}

For all datasets, we employ the word2vec algorithm \cite{mikolov2013efficient} to obtain the 128-dimensional word embeddings of each word in the text documents. Then, for each node, we average the word embedding vectors of all the words in its document, and the averaged vector is used as the node's input features for the GNN-based methods. 

\begin{table}[t]
    \small
	\centering  
	\caption{Statistics of datasets.} 
        \vspace{-2mm}
	\label{table.datasets}  
	\resizebox{1\columnwidth}{!}{
	\begin{tabular}{l|rrrr}  
		\toprule  
		Dataset  & Cora    & Art       & Industrial & M.I. \\  \midrule
        \# Documents & 25,120  & 1,615,902 & 1,260,053  & 905,453        \\
        \# Links & 182,280 & 4,898,218 & 3,101,670  & 2,692,734     \\
        \# Avg.~doc length &141.26&54.23&52.15&84.66\\
        \# Avg.~node deg &7.26&3.03&2.46&2.97\\
        \# Total classes &70&3,347&2,462&1,191\\

		\bottomrule
	\end{tabular}
	}
\end{table}

\stitle{Task construction.}
We perform zero- or few-shot text classification.
We adopt a \emph{5-way}  setting, \ie, we sample five classes from all the classes to construct a task. In each task, we construct a $K$-shot support set by further sampling $K$ examples from each class for $K \in \{0, 1,\ldots, 5\}$, and a validation set of the same size as the support set. The remaining examples form the query set.
Note that the support set is labeled and serves as task training data, whereas the query set is unlabeled and used for evaluation. Note that in our experiments, all the classes are used---it is only that each task involves five classes, and we have multiple tasks during testing to cover all the classes. This is a typical task setup \cite{finn2017model}, which allows for a comprehensive evaluation under different class combinations. The reported results are averaged over all the tasks on each dataset. 


\begin{table*}[tbp]
     \small 
	\centering 
 	\addtolength{\tabcolsep}{2pt}
    \renewcommand{\arraystretch}{1.08}
	\caption{\emph{Five-shot} classification performance (percent) with 95\% confidence intervals.  
	} 
	\label{table:few-shot} 
    {
    \vspace{-3mm}
    \footnotesize In each column, the best result among all methods is \textbf{bolded} and the best among the baselines is \underline{underlined}. Improvement by \model\ is calculated\\ relative to the best baseline.  $^{*}$ indicates that our model significantly outperforms the best baseline  based on the two-tail $t$-test $(p<0.05)$.
    }
    \\[2mm] 
	\begin{tabular}{c|cc|cc|cc|cc}  
		\toprule
		  &\multicolumn{2}{c|}{Cora}&\multicolumn{2}{c|}{Art}&\multicolumn{2}{c|}{Industrial}&\multicolumn{2}{c}{M.I.}\\\cmidrule{2-9}
		  & Accuracy&Macro-F1&Accuracy&Macro-F1&Accuracy&Macro-F1&Accuracy&Macro-F1
		 \\\midrule
		 GCN &41.15$\pm$2.41 &34.50$\pm$2.23 &22.47$\pm$1.78 &15.45$\pm$1.14 
   &21.08$\pm$0.45 &15.23$\pm$0.29 &22.54$\pm$0.82 &16.26$\pm$0.72  \\
		 SAGE$_\text{sup}$  &41.42$\pm$2.90 &35.14$\pm$2.14 &22.60$\pm$0.56 &16.01$\pm$0.28 
   &20.74$\pm$0.91 &15.31$\pm$0.37 &22.14$\pm$0.80 &16.69$\pm$0.62  \\
        TextGCN &59.78$\pm$1.88 &55.85$\pm$1.50 &43.47$\pm$1.02 &32.20$\pm$1.30 
        &53.60$\pm$0.70 &45.97$\pm$0.49 &46.26$\pm$0.91 &38.75$\pm$0.78  \\
		 \midrule
		 GPT-GNN &76.72$\pm$2.02 &72.23$\pm$1.17 &65.15$\pm$1.37 &52.79$\pm$0.83 
   &62.13$\pm$0.65 &54.47$\pm$0.67 &67.97$\pm$2.49 &59.89$\pm$2.51  \\
		 DGI &\underline{78.42}$\pm$1.39 &\underline{74.58}$\pm$1.24 &65.41$\pm$0.86 &53.57$\pm$0.75 
   &52.29$\pm$0.66 &45.26$\pm$0.51 &68.06$\pm$0.73 &60.64$\pm$0.61  \\
		 SAGE$_\text{self}$ &77.59$\pm$1.71 &73.47$\pm$1.53 &76.13$\pm$0.94 &65.25$\pm$0.31 
   &71.87$\pm$0.61 &65.09$\pm$0.47 &\underline{77.70}$\pm$0.48 &\underline{70.87}$\pm$0.59  \\
		 \midrule
        BERT &37.86$\pm$5.31 &32.78$\pm$5.01 &46.39$\pm$1.05 &37.07$\pm$ 0.68
        &54.00$\pm$0.20 &47.57$\pm$0.50 &50.14$\pm$0.68 &42.96$\pm$1.02  \\
        BERT$^{*}$ &27.22$\pm$1.22 &23.34$\pm$1.11 &45.31$\pm$0.96 &36.28$\pm$0.71  
        &49.60$\pm$0.27 &43.36$\pm$0.27 &40.19$\pm$0.74 &33.69$\pm$0.72  \\
        RoBERTa &62.10$\pm$2.77 &57.21$\pm$2.51 &72.95$\pm$1.75 &62.25$\pm$1.33 
        &76.35$\pm$0.65 &70.49$\pm$0.59 &70.67$\pm$0.87 &63.50$\pm$1.11  \\
        RoBERTa$^{*}$ &67.42$\pm$4.35 &62.72$\pm$3.02 &74.47$\pm$1.00 &63.35$\pm$1.09 
        &77.08$\pm$1.02 &71.44$\pm$0.87 &74.61$\pm$1.08 &67.78$\pm$0.95  \\
		
		\midrule
        P-Tuning v2 &71.00$\pm$2.03 &66.76$\pm$1.95 &\underline{76.86}$\pm$0.59 &\underline{66.89}$\pm$1.14 
        &\underline{79.65}$\pm$0.38 &\underline{74.33}$\pm$0.37 &72.08$\pm$0.51 &65.44$\pm$0.63  \\
		
		\midrule
		\model-p &79.16$\pm$1.23 &74.99$\pm$1.35 &79.59$\pm$0.31 &68.26$\pm$0.43 
  &80.86$\pm$0.40 &74.44$\pm$0.29 &81.26$\pm$0.36 &74.82$\pm$0.45  \\
		\model\ &\textbf{80.08}$^{*}${}$\pm$1.33 &\textbf{75.91}$^{*}${}$\pm$1.39 
        &\textbf{81.03}$^{*}${}$\pm$0.43 &\textbf{69.86}$^{*}${}$\pm$0.67 
        &\textbf{82.46}$^{*}${}$\pm$0.29 &\textbf{76.36}$^{*}${}$\pm$0.25 
        &\textbf{82.77}$^{*}${}$\pm$0.32 &\textbf{76.48}$^{*}${}$\pm$0.52  \\
		(improv.) & (+2.12\%)&(+1.78\%) & (+5.43\%)&(+4.44\%) 
  & (+3.53\%)&(+2.7\%) & (+6.53\%)&(+7.92\%)\\
	\bottomrule
	\end{tabular}
\end{table*}

\stitle{Baselines for few-shot classification.}
We consider competitive baselines from four categories. 

(1) \emph{End-to-end GNNs}, which are graph neural networks trained in a supervised, end-to-end manner from random initialization. \vspace{-.5mm}
\begin{itemize}[leftmargin=*]
    \item GCN \cite{KipfW17}: an extension of the convolutional neural network that operates on the graph. 
    \item SAGE$_\text{sup}$ \cite{hamilton2017inductive}: the supervised version of GraphSAGE, an inductive GNN that generates node embeddings by sampling and aggregating features from a node's local neighborhood.
    \item TextGCN \cite{yao2019graph}: a GCN-based model on a text graph constructed from word co-occurrence and document-word relations, which jointly learns the embeddings of both words and documents.
\end{itemize}
(2) \emph{Pre-trained/self-supervised GNNs}, these GNNs are pre-trained using pretext tasks without labeled data, followed by fine-tuning or  fitting a classification head while freezing the model parameters. \vspace{-.5mm}
\begin{itemize}[leftmargin=*]
    \item GPT-GNN \cite{hu2020gpt}: a GNN pre-training approach by a self-supervised graph generation task, including node attribute generation and edge generation. It follows the ``pre-train, fine-tune" paradigm.
    \item DGI \cite{velickovic2019deep}: a GNN pre-training approach that maximizes the mutual information between local and global representations. As an unsupervised method, it also  freezes the model parameters and fits a simple logistic regression model for the downstream few-shot classification, after pre-training.
    \item SAGE$_\text{self}$ \cite{hamilton2017inductive}: the self-supervised version of GraphSAGE, encouraging similar embeddings for neighboring nodes and distinct embeddings for non-adjacent nodes. After pre-training, it follows the same approach of DGI for the downstream classification.
\end{itemize}
(3) \emph{Pre-trained transformers}, which are pre-trained using masked language modeling \cite{kenton2019bert}, and then fine-tuned together with a randomly initialized classification head (\eg, a fully connected layer), for the downstream few-shot classification task. \vspace{-.5mm}
\begin{itemize}[leftmargin=*]
    \item BERT \cite{kenton2019bert}: a bidirectionally trained transformer using masked language modeling, which learns from unlabeled text by being jointly conditioned on both left and right contexts in all layers.  
    \item RoBERTa \cite{liu2019roberta}: a replication of BERT that carefully measures the impact of many key hyperparameters and training data size during training. 
    \item BERT$^{*}$ and RoBERTa$^{*}$: variants of BERT and RoBERTa, which are obtained by fine-tuning the pre-trained BERT and RoBERTa, respectively, using masked language modeling  on our datasets, to mitigate the domain gap between our datasets and the datasets used for pre-training BERT and RoBERTa.
\end{itemize}
(4) \emph{Prompt tuning}: P-Tuning v2 \cite{liu-etal-2022-p}, is a version of prefix-tuning \cite{li2021prefix} optimized and adapted for natural language. It uses deep prompt tuning, which applies continuous prompts for every layer of the pre-trained language model. 

Note that our setting is different from few-shot learning under the meta-learning paradigm \cite{finn2017model}, since there are no few-shot tasks for the meta-training phase. Hence, we cannot use state-of-the-art meta-learning models for comparison. Besides, two of the baselines we compared,  DGI and SAGE$_\text{self}$, have adopted a form of linear probe which is known to be a strong few-shot learner \cite{tian2020rethinking}.

\stitle{Baselines for zero-shot classification.}
We only compare with PLMs, as all other methods require at least one shot to work. 
For each method, we use the discrete prompt \texttt{[CLASS]} (\ie, the label text alone).
We also evaluate handcrafted prompts  ``\texttt{prompt [CLASS]}'', 
where \texttt{prompt} is a sequence of tokens found by prompt engineering, and annotate the model name with ``+d''.
Essentially, we compute the similarity between the target document and the label text of each class (with or without additional tokens), and predict the most similar class following Fig.~\ref{fig:zero-shot}.

\stitle{Settings of \model\ and baselines.}
For \model, the text encoder is a transformer \cite{vaswani2017attention}. Following CLIP \cite{radford2021learning}, we use a 63M-parameter, 12-layer 512-wide model with 8 attention heads. It operates on a lower-cased byte pair encoding (BPE) representation of the texts with a 49,152 vocabulary size \cite{sennrich2016neural}. The maximum sequence length is capped at 128. The graph encoder employs a GCN \cite{KipfW17}, using two layers \cite{hamilton2017inductive} with LeakyReLU activation, each with 128 dimensions \cite{perozzi2014deepwalk}.
The pre-training of our model starts from scratch without initializing the graph and text encoders with previously pre-trained weights.
$\lambda$ in Eq.~\eqref{eq:final loss} is set to $0.1$ on Cora, and set to $10$ on the three Amazon review datasets, which were chosen from \{0.01, 0.1, 1, 10, 100\} according to the accuracy on the validation data. The number of learnable prompt tokens, $M$ in Eq.~\eqref{eq:prompt embeddings}, is set to $4$, which was chosen from \{2, 4, 8, 16, 32\} based on the validation data. We use the Adam optimizer with the learning rate $2\times 10^{-5}$ with 2 training epochs, and a batch size of 64 in pre-training, referring to Hugging Face's \cite{wolf2020transformers} example settings. The text embedding size is 128, the same as the output of the graph encoder.
To generate the summary embedding and the context-based prompt initialization, the number of neighboring nodes sampled
is 3.
For prompt tuning, we set the learning rate as $0.01$, which was chosen from \{0.0001,0.001,0.01,0.1\} according to the accuracy on validation data.   

For all the GNN methods, including the GNN component in \model, we use the 128-dimensional word2vec embeddings \cite{mikolov2013efficient} averaged over the words in the raw texts as the input node features. We use a two-layer architecture, and set the hidden dimension to be 128, except for GCN and SAGE$_\text{sup}$ whose hidden dimension is set to 32 \cite{KipfW17} which gives better empirical performance. For all GNN pre-training baselines, we use 0.01 as the learning rate. For BERT, RoBERTa and \model, we adopt 0.00002 as the learning rate. Our implementations of BERT, RoBERTa and their masked language modeling are based on Hugging Face's transformers \cite{wolf2020transformers}. For both BERT and RoBERTa, we use their base versions, given that our model \model\ uses just a small 63M-parameter model, following previous work \cite{radford2021learning}. 
For P-Tuning v2, we use the original code on the RoBERTa backbone, and take the recommended 0.005 as the learning rate for prompt tuning. For \model, the learning rate for prompt tuning is set to 0.01.

We conduct all experiments on a server with 4 units of GeForce RTX 3090 GPU. Pre-training \model\ takes about 0.5/6/9/10 hours on Cora/M.I./Industrial/Art, respectively, on a single GPU. The inference (with prompt-tuning) is carried out with five different splits generated from five random seeds \{1, 2, 4, 8, 16\}.  


\subsection{Performance of low-resource classification}

We evaluate the classification performance under various shots.

\stitle{Five shots.} In Tab.~\ref{table:few-shot}, we first compare the performance of \model\ with baselines under the \emph{5-shot} setting. \model\ emerges as the winner consistently, outperforming the best baseline by around 2--8\% with statistical significance.

We also make a few more observations.
Firstly, among the GNNs, pre-trained/self-supervised models tend to perform better than the end-to-end approaches, since the latter heavily rely on labeled data. 
Among the former, DGI and SAGE$_\text{self}$ perform better as they are a form of linear probe, known to be a strong few-shot learner \cite{tian2020rethinking}.
Note that, instead of using word2vec embeddings \cite{mikolov2013efficient} of raw texts as node features, we also tired using the pre-trained RoBERTa \cite{liu2019roberta} to generate the node features for DGI and SAGE$_\text{self}$.
However, doing so does not bring any improvement, showing that it is ineffective to simply combine a language model and GNN in a decoupled manner. In contrast, our proposed model jointly learns the text and graph encoders through three graph-grounded contrastive strategies. 
%
Secondly, PLMs are generally superior to GNNs, illustrating the importance of leveraging texts in a fine-grained way. Additionally, RoBERTa outperforms BERT owing to an improved pre-training procedure \cite{liu2019roberta}. However, further fine-tuning PLMs on our text data gives mixed results: 
RoBERTa$^*$ slightly outperforms RoBERTa but BERT$^*$ is much worse than BERT. In other words, it is not straightforward to mitigate the domain gap by simply fine-tuning with the domain texts.
%
Thirdly,  the continuous prompt approach P-Tuning v2 achieves competitive results compared to fine-tuning, while having the advantage of being much cheaper than fine-tuning. 
However, our model \model\ still significantly outperforms it. 
Furthermore, \model-p without prompt tuning is inferior to \model, showing the benefit of continuous prompts.

\begin{figure}
   \subfigure[\# of shots, Cora]{
   \includegraphics[width=0.48\linewidth]{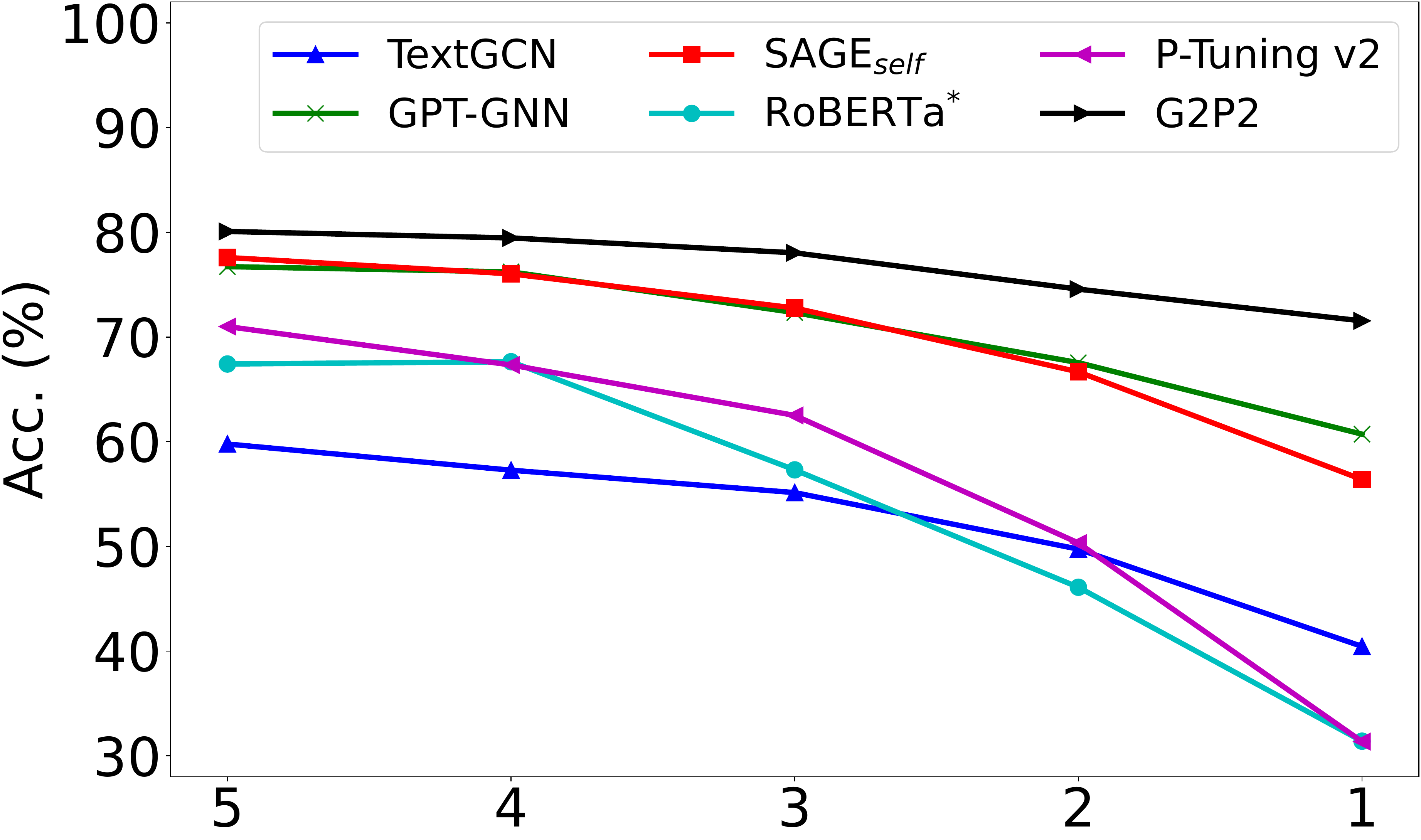}\vspace{-2mm}
   }%
   \subfigure[\# of shots, Art]{
   \includegraphics[width=0.48\linewidth]{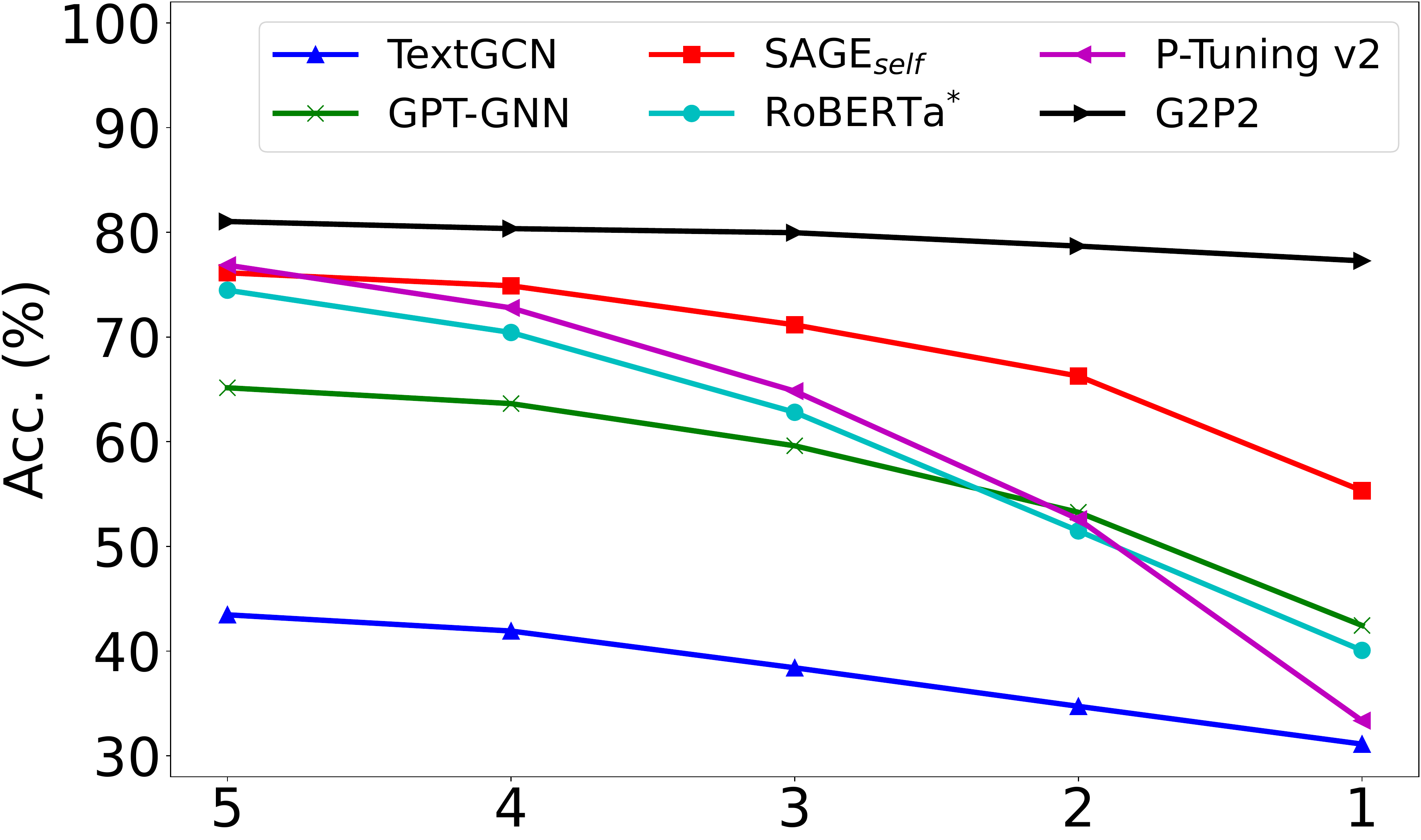}}\vspace{-2mm}
   \subfigure[\# of shots, Industrial]{
   \hspace{0.5mm}%
   \includegraphics[width=0.48\linewidth]{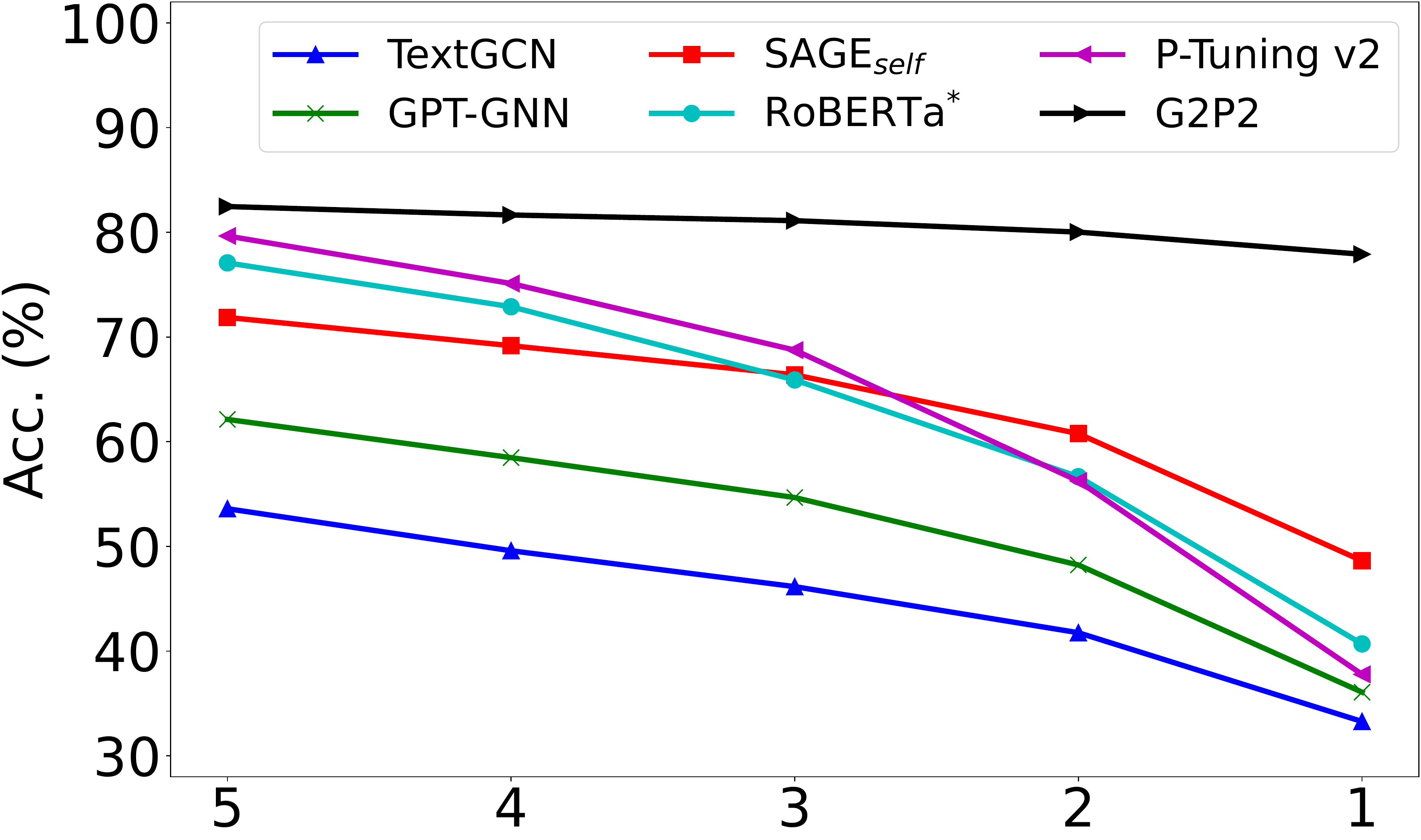}}%
   \subfigure[\# of shots, M.I.]{
   \hspace{1mm}%
   \includegraphics[width=0.48\linewidth]{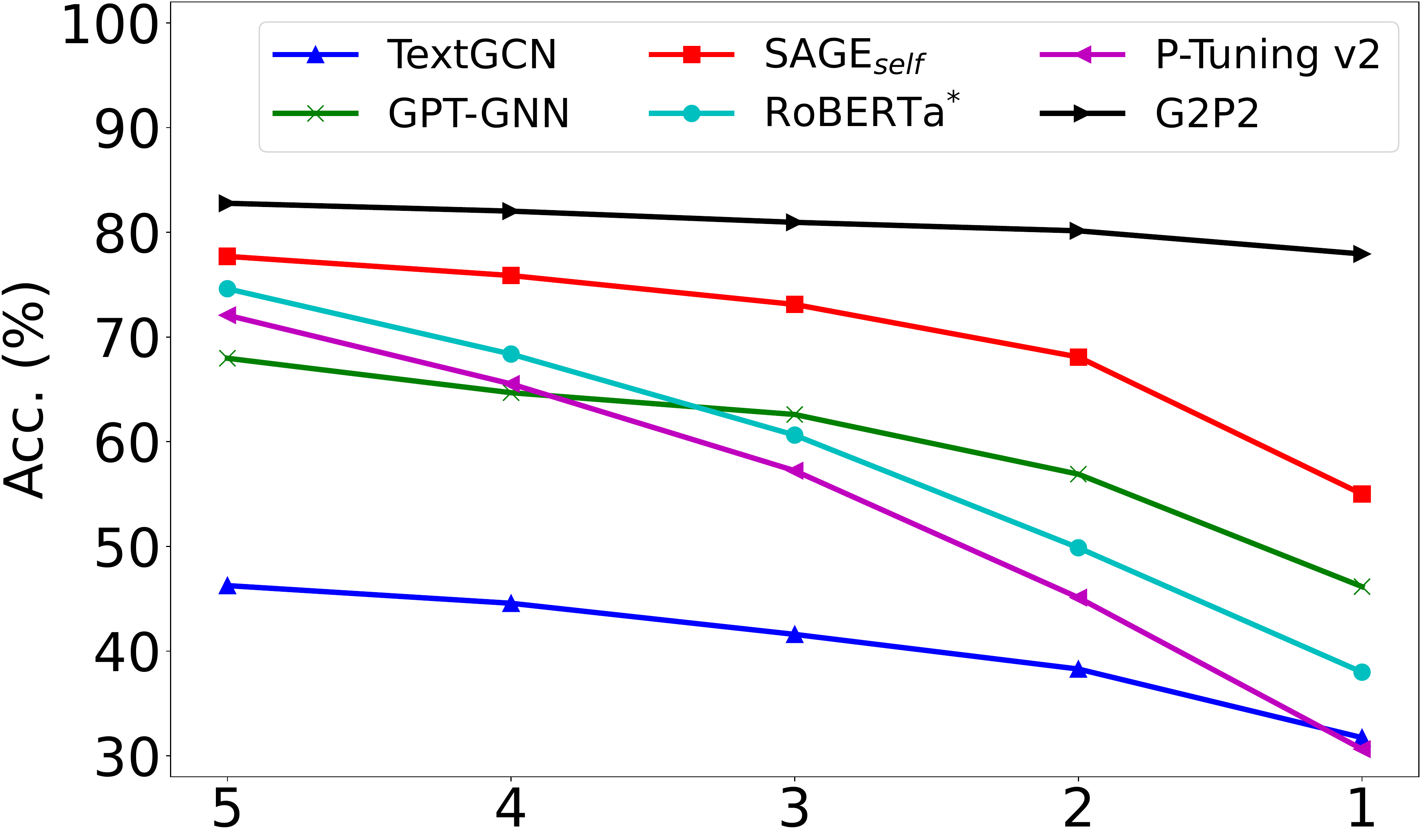}
   }
   \vspace{-3mm}
	\caption{Performance on different shots.}
	\label{fig:shots}
\end{figure}

\stitle{Fewer shots.}
In addition to the 5-shot setting, in Fig.~\ref{fig:shots} we also study the impact of fewer shots  on \model\ and several representative baselines. \model\ generally performs the best across different shots. 
In general, the performance of all approaches decreases as the number of shots is reduced. 
However, the baselines suffer significantly under extreme low-resource (\eg, 1- or 2-shot) settings. In contrast, \model\ remains robust, reporting a relatively small decrease in performance even with just 1 or 2 shots.

The results demonstrate the practical value of our proposed model especially when labeled data are difficult or costly to obtain in time. On the other hand, 
traditional approaches constantly face the challenge of the inability to keep up with the rapid growth of emerging classes in dynamic and open environments \cite{wang2021zero}.  For example, labeling a large volume of texts for novel topics in online articles, or new product categories in open-ended e-commerce platforms, can suffer a substantial time lag.


\stitle{Zero shot.}
Finally, we report the zero-shot performance in Tab.~\ref{table:zero-shot}, where our models \model\ and \model+d significantly outperform the baselines. The results particularly demonstrate the effectiveness of our graph-grounded contrastive pre-training in the absence of labeled data, which is crucial to handling evolving classes without any labeled sample in many real-world scenarios. 
Moreover, handcrafted discrete prompts (\ie, BERT$^*$+d and \model+d) can be superior to using label text only (\ie, BERT$^*$ and \model), showing the effectiveness of additional prompt tokens. 

However, finding the optimal discrete prompts often requires significant engineering work. Specifically, for the three approaches with discrete prompts, namely, RoBERTa$^{*}$+d, BERT$^{*}$+d and \model+d, we explored more than 10 handcrafted prompt templates on each dataset, which are typically relevant to the corresponding dataset and require some domain knowledge to devise.
While discrete prompts are generally helpful to zero-shot classification, their effectiveness varies. In Tab.~\ref{table:zero-shot}, we simply report the performance of the best handcrafted template for each approach and each dataset.
It is also worth noting that the same prompt can sometimes generate opposite results on different models. 
For instance, in Cora dataset, while ``\texttt{a model of [CLASS]}" is the best prompt for RoBERTa$^{*}$+d, it is a bad choice for \model+d. 
Moreover, some prompts without any semantic meaning, like ``\texttt{a [CLASS]}", can be the best choice sometimes.
The observations imply that prompt engineering involves labor-intensive work, and the outcomes contain much uncertainty on what the optimal discrete prompt would be. 
Therefore, using only the label text is still a reasonably good choice.



\begin{table}[tbp]
    \small 
	\centering 
 	\addtolength{\tabcolsep}{-1pt}
	\caption{\emph{Zero-shot} classification accuracy (percent).  
	} 
	\label{table:zero-shot} 
   {\vspace{-3mm}\footnotesize
   See Table~\ref{table:few-shot} for explanations on entry styles.
   }
    \\[2mm] 
	\begin{tabular}{c|c|c|c|c}  
		\toprule
		  &Cora&Art&Industrial&M.I.
		 \\\midrule
		 RoBERTa &30.46$\pm$2.01 &42.80$\pm$0.94 &42.89$\pm$0.97 &36.40$\pm$1.20 \\
         RoBERTa$^{*}$ &39.58$\pm$1.26 &34.77$\pm$0.65 &37.78$\pm$0.32 &32.17$\pm$0.68 \\
        RoBERTa$^{*}$+d &\underline{45.53}$\pm$1.33 &36.11$\pm$0.66 &39.40$\pm$1.22 &37.65$\pm$0.33 \\
		BERT &23.58$\pm$1.88 &35.88$\pm$1.44 &37.32$\pm$0.85 &37.42$\pm$0.80 \\
        BERT$^{*}$ &23.38$\pm$1.96 &54.27$\pm$1.85 &\underline{56.02}$\pm$1.22 &50.19$\pm$0.72 \\
        BERT$^{*}$+d &26.65$\pm$1.71 &\underline{56.61}$\pm$1.76 &55.93$\pm$0.96 &\underline{52.13}$\pm$0.88 \\
		\midrule
		\model &63.52$\pm$2.89 &76.52$\pm$0.59 &76.66$\pm$0.31 &74.60$\pm$0.62 \\
		\model+d & \textbf{65.28}$^{*}${}$\pm$3.12   & \textbf{76.99}$^{*}${}$\pm$0.60 
   & \textbf{77.43}$^{*}${}$\pm$0.27 & \textbf{75.86}$^{*}${}$\pm$0.69\\
		(improv.) & (+45.38\%)& (+36.00\%)&(+38.22\%)& (+45.52\%)\\
	\bottomrule
	\end{tabular}
\end{table}

\subsection{Model analyses}

We conduct more in-depth studies on \model. Unless otherwise stated, we report the classification \emph{accuracy} under the \emph{5-shot} setting.

\stitle{Ablation study.}
We first evaluate the contribution from each of the three graph interaction-based contrastive strategies, by employing different combinations of the proposed loss terms $\bL_1,\bL_2$ and $\bL_3$. 
As shown in Tab.~\ref{table:ablation}, strategies without $\bL_1$ have performed quite poorly, demonstrating that the bijective text-node interaction is the fundamental component of our pre-training. 
That being said, when further adding $\bL_2$ or $\bL_3$ to $\bL_1$, we still observe a noticeable performance improvement, showing the benefit of incorporating additional graph-based interactions for text data.
Lastly, \model\ with all three loss terms outperforms all 1- or 2-combinations of the losses, demonstrating that the three contrastive strategies are all useful and they are well integrated.
Overall, the results reveal that graph information is vital to low-resource text classification, since graph structures reveal rich relationships between documents.


Next, we evaluate the contribution from our prompt-tuning approach. Specifically, we compare \model\ with two ablated variants: using label text only without trainable prompt vectors, and randomly initializing the prompt vectors. As reported in Tab.~\ref{table:ablation}, only using label text clearly degrades classification performance, implying the importance of learning continuous prompts through prompt-tuning. Furthermore, our approach \model\ with context-based initialization for prompt vectors shows a small but consistent advantage over random initialization, which implies the usefulness of considering graph structures in prompt-tuning.

\begin{table}[tbp]
    \small 
	\centering 
 	\addtolength{\tabcolsep}{-0.5pt}
	\caption{Ablation study.  
	} 
	\label{table:ablation} 
 \vspace{-2mm}
	\begin{tabular}{c|c|c|c|c}  
		\toprule
		  &Cora&Art&Industrial&M.I.
		 \\\midrule
		 Only $\bL_3$  &74.66$\pm$1.80 &52.56$\pm$1.09 &45.97$\pm$0.81 &49.05$\pm$0.54\\
          Only $\bL_2$ &77.01$\pm$1.30 &58.90$\pm$0.55 &52.99$\pm$0.46 &59.41$\pm$0.85\\
          Only $\bL_1$ &79.50$\pm$1.19 &77.37$\pm$0.72 &78.10$\pm$0.34 &79.70$\pm$0.56\\
          $\bL_2$+$\bL_3$ &70.04$\pm$2.89 &49.91$\pm$1.57 &50.07$\pm$0.50 &56.14$\pm$1.01\\
          $\bL_1$+$\bL_3$ &79.73$\pm$0.89 &78.60$\pm$0.40 &79.97$\pm$0.43 &80.42$\pm$0.45\\
          $\bL_1$+$\bL_2$ &79.42$\pm$1.04 &80.55$\pm$0.52 &81.06$\pm$0.33 &82.39$\pm$0.41\\
          \midrule
          Only label text &79.16$\pm$1.23 &79.59$\pm$0.31 &80.86$\pm$0.40 &81.26$\pm$0.36\\
          Random init. &80.03$\pm$0.99 &80.85$\pm$0.43 &82.43$\pm$0.33 &82.64$\pm$0.21\\
		\midrule
		\model & \textbf{80.08}$\pm$1.33   & \textbf{81.03}$\pm$0.43  & \textbf{82.46}$\pm$0.35 & \textbf{82.77}$\pm$0.32 \\
	\bottomrule
	\end{tabular}
\end{table}

\begin{figure}
   \subfigure[Interaction coefficient, $\lambda$]{
   \centering
   \includegraphics[width=0.48\linewidth]{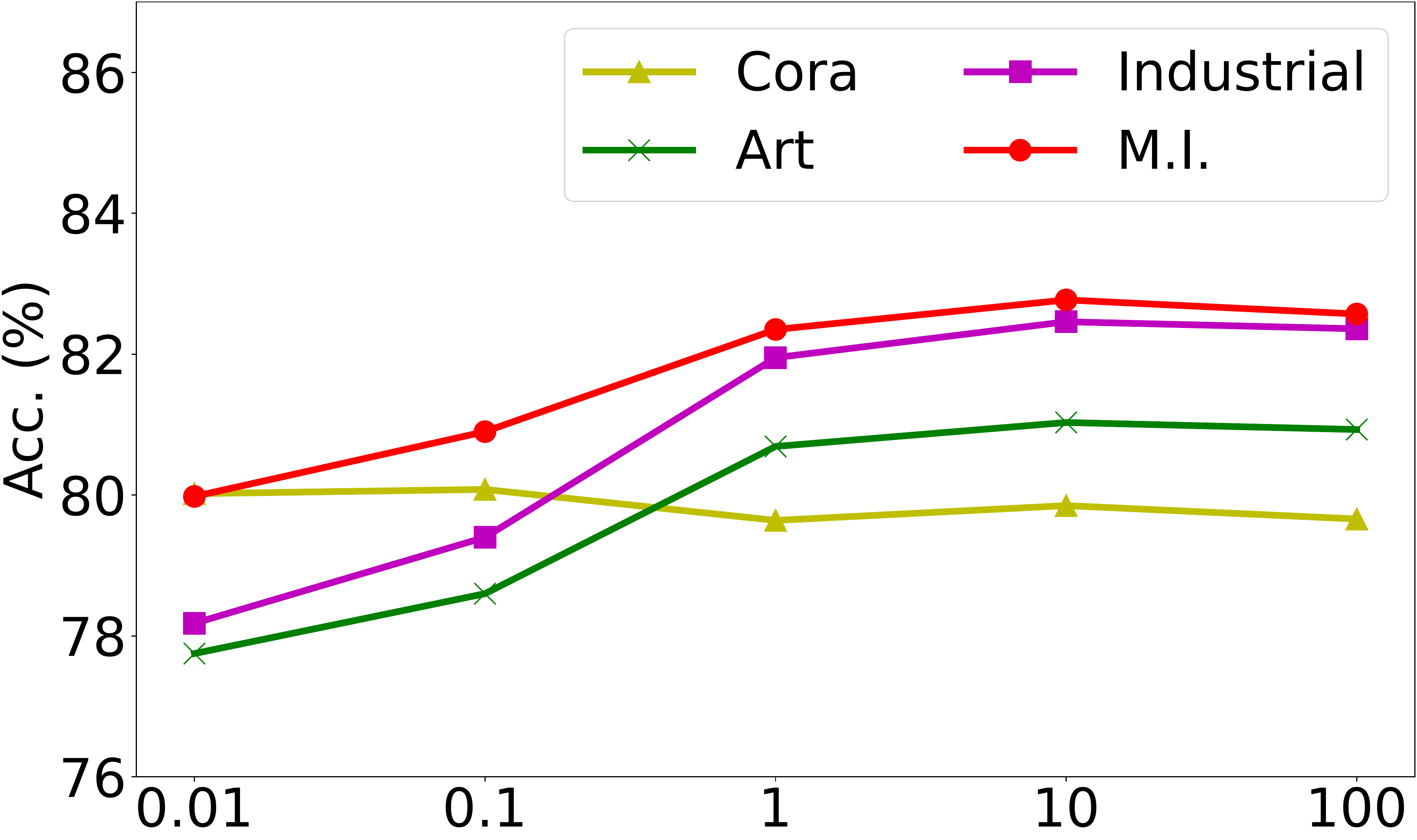}
   }
   \subfigure[Prompt length, $M$]{
   \centering
   \includegraphics[width=0.48\linewidth]{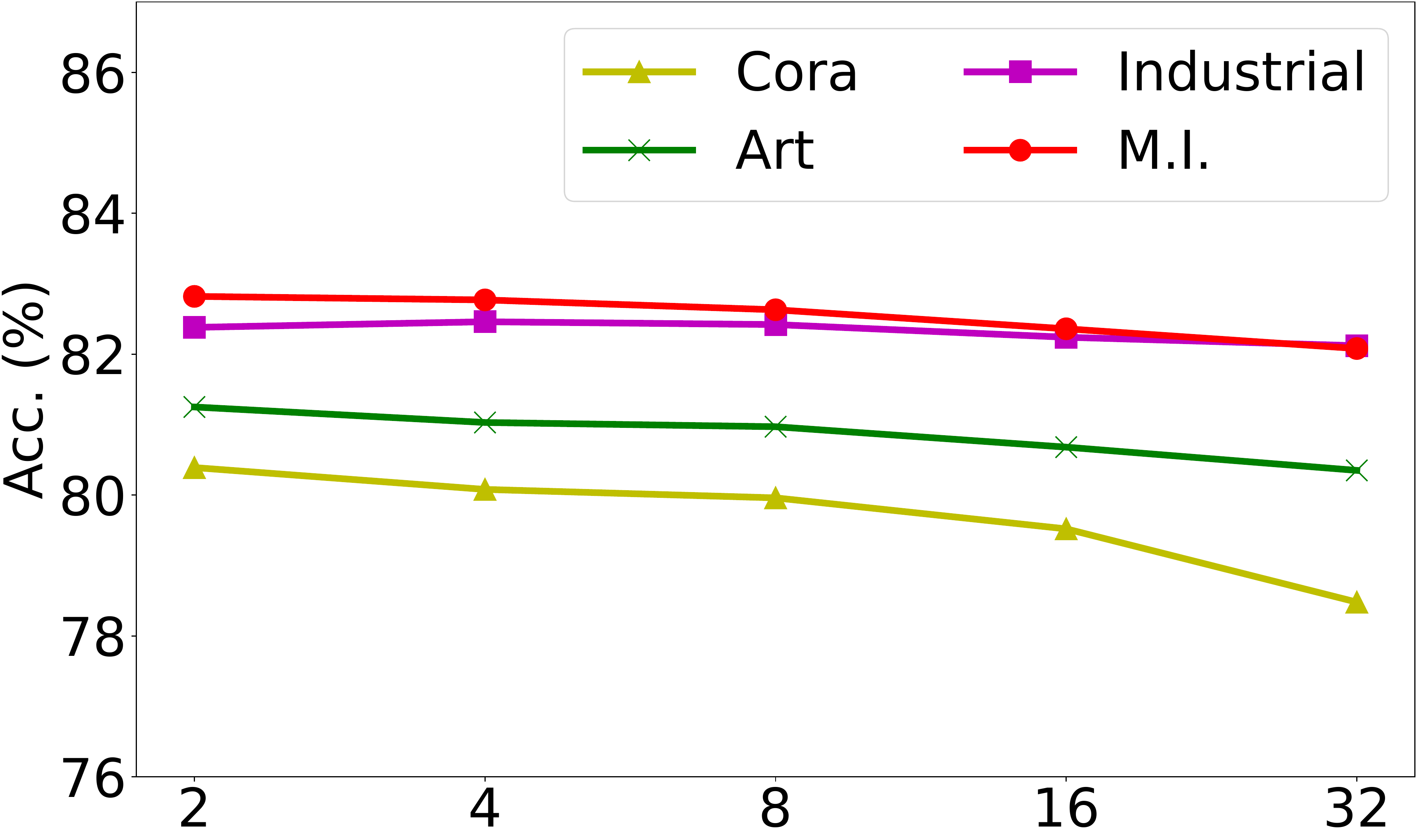}
   }
   \vspace{-2mm}
	\caption{Hyperparameter study.}
	\label{fig:param}
\end{figure}

\stitle{Hyperparameter study.}
We first investigate the impact of the interaction coefficient $\lambda$ in Fig.~\ref{fig:param}(a), which balances the high-order contrastive losses ($\bL_2,\bL_3$). The performance is generally better and stable when $\lambda$ is slightly bigger (\eg, $\ge 10$), indicating the significance of the high-order text-summary and node-summary interactions. 
Next, we study the prompt length $M$ in Fig.~\ref{fig:param}(b), which refers to the number of trainable prompt vectors in Sect.~\ref{sec:model_prompt}.
The performance is relatively unaffected by the prompt length, and thus it is robust to choose a small $M$ (\eg, 4) for efficiency.

\stitle{Efficiency of prompt tuning.}
In our work, the continuous prompts are optimized by prompt tuning \cite{liu-etal-2022-p, zhou2022learning} without updating the pre-trained model. In this experiment, we investigate the efficiency of prompt-tuning in \model\ compared to the efficiency of traditional fine-tuning. 
As \model\ has a transformer component, we compare it with four transformer based models, all of which follow the classical ``pre-train, fine-tune'' paradigm \cite{kenton2019bert}.

As shown in Tab.~\ref{table:efficiency}, ``Tuning time per task" refers to the average time required per task for prompt-tuning in \model\ or fine-tuning in the baselines, while ``Param.~size" refers to the number of parameters that require updating.
The results demonstrate that prompt tuning in \model\ is much more efficient than fine-tuning in the baselines, achieving 2.1$\sim$18.8x speedups. The reason is that prompt tuning updates far fewer parameters. In \model, we used 4 trainable 512-dimensional prompt vectors, totaling 2048 parameters only, while fine-tuning in the baselines needs to update the whole pre-trained model with more than 100M parameters. Note that the speedup is not linear w.r.t.~the parameter size, due to the overhead in the data loader and the optimizer. 
Overall, our prompt tuning is not only effective under low-resource settings, but also parameter- and computation-efficient.
\begin{table}[tbp]
    \small 
	\centering 
 	\addtolength{\tabcolsep}{-1.5pt}
	\caption{Tuning time and parameter size. } 
	\label{table:efficiency} 
 \vspace{-2mm}
	\begin{tabular}{c|c|c|c|c|c}  
		\toprule
       & \multicolumn{4}{c|}{Tuning time per task (in seconds)} & Param. \\\cline{2-5}
		  &Cora&Art&Industrial&M.I.&size
		 \\\midrule
		 RoBERTa & 45.47$\pm$2.38 & 64.22$\pm$3.62 & 43.46$\pm$2.99 & 44.99$\pm$2.58 & 123 M\\
          RoBERTa$^{*}$ & 39.38$\pm$2.01 & 59.56$\pm$3.55 & 35.10$\pm$2.75 & 38.84$\pm$2.39 & 123 M\\
		 BERT & 32.23$\pm$1.71 & 51.77$\pm$2.00 & 31.72$\pm$1.77 & 33.55$\pm$2.39 & 110 M\\
          BERT$^{*}$ & 34.82$\pm$1.68 & 55.16$\pm$2.32 & 31.11$\pm$1.74 & 29.00$\pm$2.23 & 110 M\\
		\midrule
		\model & \textbf{2.42}$\pm$0.41   & \textbf{22.03}$\pm$1.39 & \textbf{14.63}$\pm$1.26 & \textbf{12.72}$\pm$1.17 & \textbf{2048}\\
	\bottomrule
	\end{tabular}
\end{table}

\stitle{Generalization study.}
Our previous experiments can be considered ``transductive'' as both the pre-training of the text encoder and the downstream classification are conducted on the whole corpus.
To further evaluate the generalization ability of our model, we adopt an ``inductive'' setting, whereby we pre-train the text encoder only on a subset of the corpus and perform downstream classification on a disjoint subset. Particularly, in the three Amazon datasets, since user texts have no labels and item texts have labels, it is natural for us to pre-train with only user texts and classify only item texts downstream. We also employ masked language modeling on only the user texts for BERT and RoBERTa, to get BERT$^{*}$ and RoBERTa$^{*}$. As shown in Tab.~\ref{table:generalization}, \model\ still performs very well in the inductive setting, illustrating the strong generalization ability of our pre-trained model.

\begin{table}[tbp]
    \small 
	\centering 
 	\addtolength{\tabcolsep}{6pt}
	\caption{Inductive performance on text classification.} 
	\label{table:generalization} 
 \vspace{-2mm}
	\begin{tabular}{c|c|c|c}  
		\toprule
		  &Art&Industrial&M.I.
		 \\\midrule
          BERT$^{*}$ &43.66$\pm$0.90  &48.35$\pm$0.25 &39.24$\pm$0.88 \\
		 RoBERTa$^{*}$ &69.55$\pm$1.14  &73.65$\pm$0.86 &71.96$\pm$1.44 \\ 
		\midrule
		\model& \textbf{79.81}$\pm$0.22   & \textbf{81.29}$\pm$0.32
  & \textbf{81.85}$\pm$0.33  \\
	\bottomrule
	\end{tabular}
\end{table}

\section{Conclusion}
In this paper, we studied the problem of low-resource text classification. Given that many text documents are related through an underlying network, we proposed a novel model called Graph-Grounded Pre-training and Prompting (\model). It consists of three graph interaction-based contrastive strategies in pre-training, and a prompting mechanism for the jointly pre-trained graph-text model in downstream classification. We conducted extensive experiments and showed the advantages of \model\ in zero- and few-shot text classification.

A limitation of this work is the need of a graph to complement the texts. Although graphs are ubiquitous in information retrieval applications, in the case that an organic graph is unavailable, a
potential solution is to construct synthetic graphs based on word co-occurrences or other relations, \eg, linking up news articles
in close time periods and locations. We leave further explorations to future work.

\begin{acks}
This research / project is supported by the Ministry of Education, Singapore, under its Academic Research Fund Tier 2 (Proposal ID: T2EP20122-0041). Any opinions, findings and conclusions or recommendations expressed in this material are those of the author(s) and do not reflect the views of the Ministry of Education, Singapore.
\end{acks}

\clearpage
\balance
\bibliographystyle{ACM-Reference-Format}
\bibliography{references.bib}

\end{document}